\documentclass[useAMS,usenatbib]{mn2e}
\usepackage{amsmath}
\usepackage{amssymb}
\usepackage{graphicx}
\usepackage{pdflscape}

\title[Transit Shapes and SOMs]{Transit Shapes and Self Organising Maps as a Tool for Ranking Planetary Candidates: Application to \emph{Kepler} and \emph{K2}}

\author[Armstrong, D. J. et. al.]{\parbox{\textwidth}{D. J. Armstrong$^{1,2}$\thanks{d.j.armstrong@warwick.ac.uk}, D. Pollacco$^1$, A. Santerne$^{3,4}$}
\vspace{0.4cm}\\
\parbox{\textwidth}{$^{1}$University of Warwick, Department of Physics, Gibbet Hill Road, Coventry, CV4 7AL, UK\\
$^{2}$ARC, School of Mathematics \& Physics, Queen's University Belfast, University Road, Belfast BT7 1NN, UK\\
$^{3}$Aix Marseille Universit\'{e}, CNRS, Laboratoire d'Astrophysique de Marseille UMR 7326, 13388, Marseille, France\\
$^{4}$Instituto de Astrof\'{i}sica e Ci\^{e}ncias do Espa\c{c}o, Universidade do Porto, CAUP, Rua das Estrelas, 4150-762 Porto, Portugal}}

\newcommand{\mytilde}{\raise.17ex\hbox{$\scriptstyle\mathtt{\sim}$}}

\begin{document}
\date{Accepted . Received}

\pagerange{\pageref{firstpage}--\pageref{lastpage}} \pubyear{2002}

\maketitle

\begin{abstract}
A crucial step in planet hunting surveys is to select the best candidates for follow up observations, given limited telescope resources. This is often performed by human `eyeballing', a time consuming and statistically awkward process. Here we present a new, fast machine learning technique to separate true planet signals from astrophysical false positives. We use Self Organising Maps (SOMs) to study the transit shapes of \emph{Kepler} and \emph{K2} known and candidate planets. We find that SOMs are capable of distinguishing known planets from known false positives with a success rate of 87.0\%, using the transit shape alone. Furthermore, they do not require any candidates to be dispositioned prior to use, meaning that they can be used early in a mission's lifetime. A method for classifying candidates using a SOM is developed, and applied to previously unclassified members of the \emph{Kepler} KOI list as well as candidates from the \emph{K2} mission. The method is extremely fast, taking minutes to run the entire KOI list on a typical laptop. We make \texttt{Python} code for performing classifications publicly available, using either new SOMs or those created in this work. The SOM technique represents a novel method for ranking planetary candidate lists, and can be used both alone or as part of a larger autovetting code.
\end{abstract}


\section{Introduction}
Transit surveys both from the ground and space have been the most successful method of discovering planets to date. Instruments such as SuperWASP\citep{Pollacco:2006im}, HAT/HATnet \citep{Bakos:2004gx}, KELT \citep{Siverd:2012em}, \emph{Kepler} \citep{Borucki:2010dn}, \emph{K2} \citep{Howell:2014ju} and \emph{CoRoT} \citep{Auvergne:2009en} have found thousands of transiting exoplanets with a wide range of parameters. The lightcurves produced by these instruments are searched for planets using techniques such as the BLS algorithm \citep{Kovacs:2002ho}. Lists of planetary candidates are produced, with some selected for further followup observations. While such lists contain many true planetary signals, they also contain instrumental signatures and astrophysical false positives such as contaminating eclipsing binaries \citep[e.g.][]{Almenara:2009fb,Santerne:2012jf,Santerne:2016fz}.

The process of selecting the best and most likely real candidates to progress to further observations is a difficult one. Typically human inspection is used to select the best candidates \citep[e.g.][]{Pope:2016kh}, a process which can be both time consuming and subject to biases. Some recent methods have been developed to address this problem \citep{McCauliff:2015fb,Coughlin:2016wa}, and we aim to present an enhancement to these here. We introduce a novel technique designed to separate planetary signals from false positives using the shape of the transit signal, utilising Self Organising Maps \citep[SOMs,][]{Kohonen:1982dy,Kohonen:1990fd,Brett:2004cr,Armstrong:2016br}. Both investigations of the transit shape \citep[e.g.][]{Thompson:2015ic} and SOMs have been used in the astrophysical literature before, but not as yet in combination. Here we apply SOMs to space-based data from the \emph{Kepler} and \emph{K2} missions, as this data is both public and provides a large number of known planets for testing. In the future we aim to explore the applicability of the technique to ground based surveys.

SOMs are a machine learning technique first introduced by \citet{Kohonen:1982dy}. They have been used in astronomy for estimating galaxy photometric redshifts \citep{CarrascoKind:2014gb}, identifying variable stars \citep{Brett:2004cr,Armstrong:2016br} and investigating active galactic nuclei \citep{Torniainen:2008cc}. Machine learning in general is a promising area only beginning to find application to the exoplanet field. Recent uses include the automatic selection of candidates using random forests \citep{McCauliff:2015fb}, and the automatic choice of planetary atmosphere components to include in models using deep neural networks \citep{Waldmann:2016dz}. \citet{Mislis:2015cj} explored the use of random forests in planet detection, but did not test their method on lightcurves showing significant out of transit variability, and concentrated on white-noise simulated lightcurves. In the current age of increasingly large surveys with previously unseen quantities of data, such automated techniques will prove necessary to fully exploit observations, for example in identifying planets, variable stars \citep{Richards:2012ea,Masci:2014bk,Armstrong:2016br} or other interesting objects. The ability to automate parts of the planetary discovery process will allow the removal of biases introduced by human intervention, making future statistical studies easier to perform and more robust.

We present this technique both as a standalone ranking method for planetary candidates, and as a potential stage in the candidate selection process, suitable for combining with more complex methods. It is simple in concept and computationally cheap. The technique is described in Section \ref{sectSOM}, and methods of using it to classify candidates detailed in Section \ref{sectClassify}. We apply the SOM to planet candidates from the \emph{Kepler} and \emph{K2} missions, demonstrating its use and ranking those candidates, in Sections \ref{sectKepler} and \ref{sectK2}. Strengths and weaknesses of the technique are discussed in Section \ref{sectDiscuss}.

\section{Data}
\label{sectData}
\subsection{\emph{Kepler}}
Data from the \emph{Kepler} satellite was used to provide a large set of already classified planets and false positives for testing. We are also able to classify currently unclassified candidates (see Table \ref{tabkepresults}). This data spans approximately 4 years, with a cadence of 1766s. Additional data with a shorter cadence near 1 minute is also available for some targets; we ignore this to maintain a uniform sample. We use the set of Kepler Objects of Interest (KOIs) available on the NASA Exoplanet Archive as of 17th May 2016. We download the Data Validation \citep[DV,][]{Wu:2010cd} lightcurves directly from the archive\footnote{http://exoplanetarchive.ipac.caltech.edu/}. These are the lightcurves used to detect the KOIs, and so we deem them the most relevant for testing this method. There were 6384 total dispositioned KOIs with DV lightcurves available. These include 2247 confirmed planets, 1785 candidates, and 2352 false positives.  Transit parameters (period, epoch, duration) are taken from the archive. 

To prepare the transit shapes for entry in to the SOM, each lightcurve is phase folded at the given ephemeris. We then cut to a region of phase within 1.5 transit durations of transit centre, making a 3 transit duration window. This region is binned down to 50 points, which we find gives a suitable resolution on the transit shape, and allows for uniform shape comparisons across KOIs with very different numbers of data points in transit. KOIs with few transits, such that any bin is left unpopulated even after phase folding, are ignored, leaving 6350 signals. Multiple planetary systems do not have other signals removed before phase folding and binning; we find that this has negligible effect on the SOM, and hence demonstrates its resilience to additional signatures both detected and undetected.

\subsection{\emph{K2}}
We also apply the method to data from \emph{Kepler}'s successor mission \emph{K2}. In this instance a substantially smaller number of candidates is presently available, due to differences between the missions and the relative youth of \emph{K2}. \emph{K2} observes single fields for \mytilde 80 day campaigns, of which 8 have been released to date. The cadence is the same as \emph{Kepler}. We use the list of candidates presented in \citet{Crossfield:2016vg}, providing 184 objects including 108 planets and 21 false positives. We used ephemeris and transit durations as provided in that work. There are many options for detrending the raw \emph{K2} data; we utilise the \texttt{EVEREST} pipeline \citep{Luger:2016ul} here, downloading data using the command line tool provided in that work.

\emph{K2} data were prepared similarly to \emph{Kepler}. Due to the shorter baseline available with \emph{K2}, fewer data points are often available for a given transit signal. The result is that 50 bins proved to be too many in several cases, leading to many targets with empty bins in the 3 transit duration window. We trialled smaller numbers of bins, finding that 20 bins was adequate for the SOM to perform. In cases where bins did not have any datapoints falling in their phase range, we linearly interpolate between nearby bin values.

\subsection{\texttt{PASTIS}}
\label{sectPASTIS}
We utilise simulated lightcurves for both testing and analysis of the SOM. These were generated with the \texttt{PASTIS} code \citep{Diaz:2014kd,Santerne:2015bb}, which produces lightcurves for various astrophysical scenarios while constraining the false positive probability of planetary candidate signals. The ability to simulate lightcurves for different scenarios allows us to test degeneracies in the SOM method. Here we create 1000 systems each for the 6 following scenarios: Planets (P), Eclipsing Binaries (EB), Eclipsing Triples (objects consisting of an eclipsing binary and companion, ET), Planets transiting the secondary star of a binary (PSB), Background Eclipsing Binaries (BEB), and Background Transiting Planets (BTP). Each system was drawn from the set of priors given in Table \ref{tabPASTISpriors}. 10000 noise-free points were simulated within one orbital period, providing a phase curve for testing or injection into real data. The phase curve was smeared to account for the \emph{Kepler} and \emph{K2} long cadence exposure time.

\begin{table}
\caption{Priors used to draw simulated systems for testing}
\label{tabPASTISpriors}
\begin{tabular}{lr}
\hline
Parameter & Prior \\
\hline
\textbf{Non Scenario Specific} &  \\
$P$ & Jeffreys 0.3d to 100d\\
$e$, $P>10$d & Beta as \citet{Kipping:2013he}\\
$e$, $P<=10$d & 0\\
$\omega$ & Uniform 0 to 360$^\circ$\\
$i$ & Uniform in sin$i$, must transit \\
\textbf{Target Star} & \\
Mass & Normal $1\pm0.15$M$_\odot$\\
$[Fe/H]$ & Normal $0\pm0.2$dex\\
age & Normal $5\pm2$Gyr$^a$\\
distance & 100pc\\
LD coefficients& \citet{Claret:2011gy}\\
\textbf{Planet} & \\
$R_p$ & Power law in $R_p^{-2}$, $1R_\oplus$ to $2.2R_\textrm{jup}$\\
$M_p$ & Normal, Expected mass $\pm50$\% \\ 
albedo & 0.1\\
\textbf{Bound Star} & \\
Mass & IMF of \citet{Kroupa:2001ki}\\
age & Fixed to target star\\
$[Fe/H]$ & Fixed to target star\\
\textbf{Background Star} & \\
Mass & IMF of \citet{Kroupa:2001ki}\\
$[Fe/H]$ & Uniform -2.5 to 0.5 dex\\
age & Uniform 0.1 to 13.7Gyr$^a$\\
distance & Power law in $D^2$, 200pc to 8kpc\\
interstellar extinction & 0.7mag/kpc$^b$\\
\hline
\multicolumn{2}{l}{$^a$ Old or massive stars excluded.}\\
\multicolumn{2}{l}{$^b$ Default in Besan\c{c}on galactic model.}
\end{tabular}
\end{table}

\section{Self Organising Map}
\label{sectSOM}
A Self Organising Map (SOM) is an \emph{unsupervised} machine learning algorithm. It finds clusters in the data given to it, without needing labels for that data to be preassigned. We will briefly describe the method here. For a more detailed overview, we refer the reader to \citet{Armstrong:2016br} or \citet{Brett:2004cr}. 

A SOM consists of an N-dimensional array of `pixels', in this case with periodic boundaries. The number of dimensions is unimportant here, but the number of pixels must be high enough to represent adequate variation in the input data for the task at hand. A good rule of thumb is to make sure you have a a few times more pieces of input data than pixels. Each pixel is a template with the same form as the input data, and is initially randomised. At the end of training, each pixel will resemble a significant pattern in the input data (a typical planetary transit or binary eclipse for example), allowing such patterns to be investigated.

`Training' the SOM occurs over a set number of iterations. In each iteration, each piece of input data (here a single transit signal) is compared to the set of SOM pixels. The best matching pixel is determined, via the minimum Euclidean distance between the input data and pixels. That pixel and those near it are altered to become slightly closer to the piece of input data under consideration. The level of change allowed is determined by the \emph{learning rate}, $\alpha$. Pixels are altered based on their proximity to the best matching pixel, determined by the \emph{learning radius} $\sigma$. Both $\alpha$ and $\sigma$ decay during the course of the training, allowing finer levels of detail to emerge.

In our case we are feeding phase folded, binned transit lightcurves into the SOM, prepared as described in Section \ref{sectData}. As such, each SOM pixel will have 50 values (20 for \emph{K2}), which form a template binned transit shape. The goal is to separate the input signals into groups of similar shape, and see if such groups have any power in distinguishing true planets from false positives. We utilise the SOM code provided in the \texttt{PyMVPA} \texttt{Python} package \citep{Hanke:2009bm}\footnote{http://www.pymvpa.org}. We use 500 training iterations. We set $\alpha=0.1$ initially, with a linear decay to zero through the course of the iterations. We set $\sigma=20$ initially (the radius of the SOM) and to decay exponentially as described in \citet{Armstrong:2016br}. The values and decay forms of $\alpha$ and $\sigma$ do not have a strong effect on the SOM's performance \citep{Brett:2004cr}.

For the \emph{Kepler} data we use a 20x20 2 dimensional SOM consisting of 400 pixels, on 6350 KOIs. As only 184 candidates are available for \emph{K2}, we reduce the size of the SOM to 8x8 and reduce $\sigma$ accordingly. We choose 2 dimensions for ease of visualisation.

\section{Classification}
\label{sectClassify}
Once training is complete, a given candidate transit shape can be placed on the SOM by finding the best matching SOM pixel. The location of this pixel, ($x$,$y$), and Euclidean distance to it, can be extracted. The challenge at this point is to classify the pixel itself, and hence the candidate under consideration; a priori, we do not know if the pixel represents a planet transit or false positive. The pixel may also be unable to distinguish between the two. Here we consider two key cases. Firstly, when a large sample of classified objects is already available (i.e. \emph{Kepler}, Case 1), and secondly earlier in a mission lifetime, when few or zero candidates have already been classified (i.e. \emph{K2}, Case 2).

\subsection{Case 1: Late in Mission Lifetime}
\label{sectCase1}
\begin{figure}
\resizebox{\hsize}{!}{\includegraphics{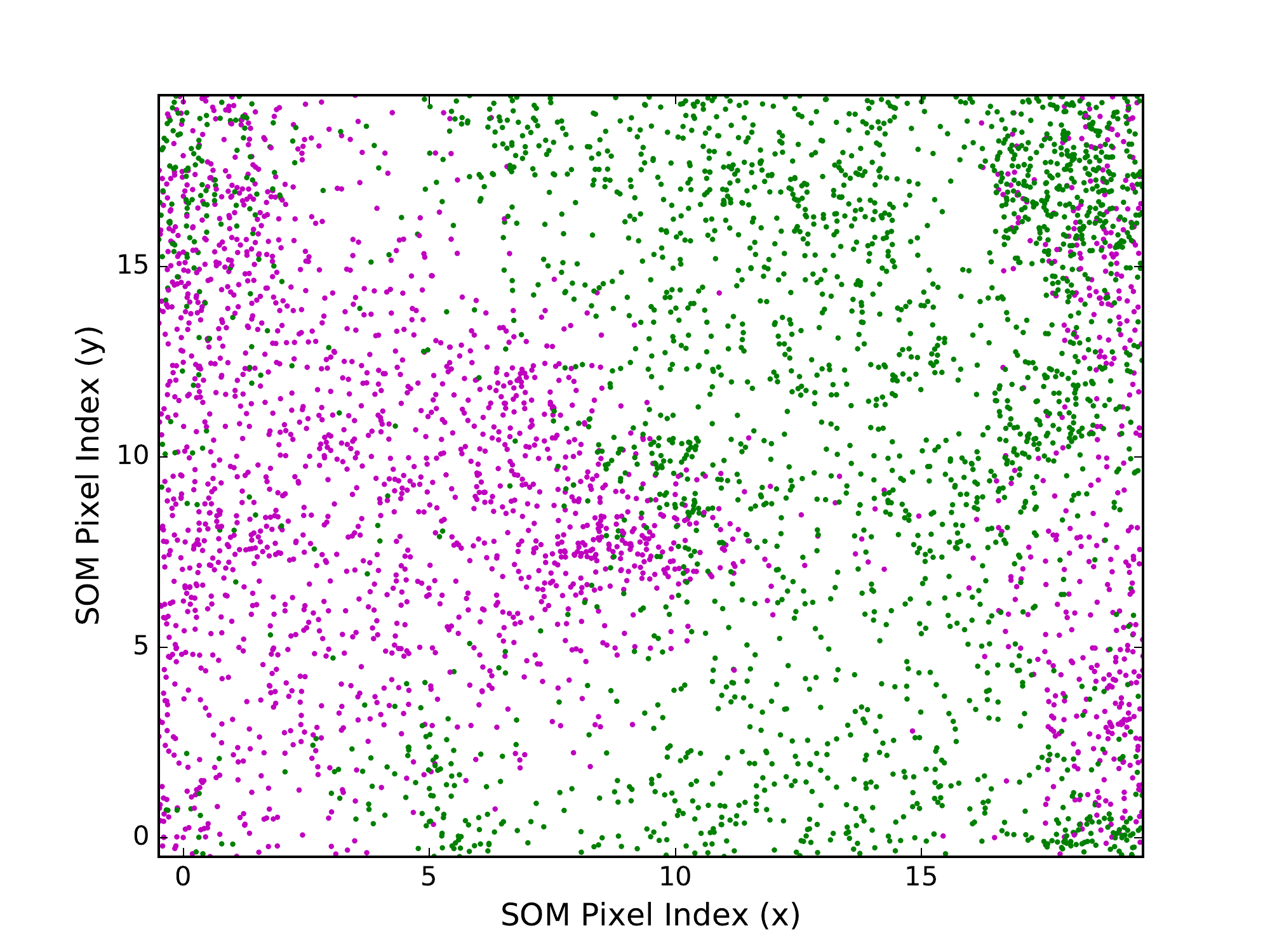}}
\caption{Trained SOM using higher SNR \emph{Kepler} KOIs. KOI signals are plotted on their best matching SOM pixel, with a random offset between -0.5 and 0.5 added to each point for clarity. Planets are magenta, false positives green. The boundaries are periodic.}
\label{figsomlockepsnrcut}
\end{figure}

\begin{figure}
\resizebox{\hsize}{!}{\includegraphics{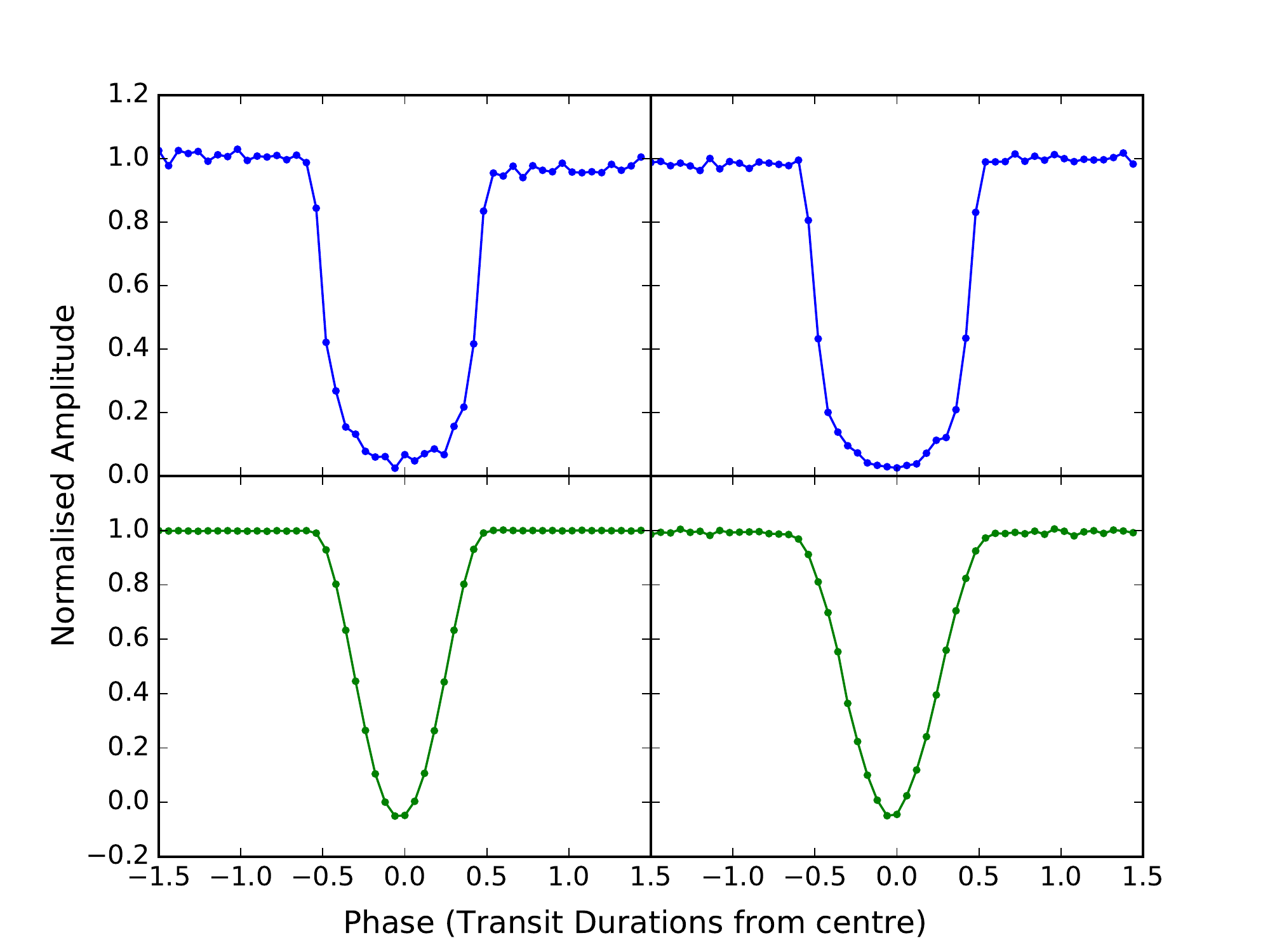}}
\caption{SOM pixel templates extracted from the SOM plotted in Figure \ref{figsomlockepsnrcut}. Clockwise from top left the templates are from pixels [1,8], [5,10], [14,3] and [11,17]. The top two templates are from regions dominated by validated planets, and the bottom two templates from regions dominated by false positives.}
\label{figtemplates}
\end{figure}

In this case, a large sample of already dispositioned signals is available. We place the sample of Kepler KOIs on the trained Kepler SOM in Figure \ref{figsomlockepsnrcut}, and show example trained SOM pixel templates in Figure \ref{figtemplates}. It is clear that the SOM has power in separating confirmed planets from false positives. Note that the key distinction is between V-shaped and U-shaped transits, something that will be discussed in Section \ref{sectDiscuss}.


To classify a candidate signal into one of the two groups, we follow the following method. Firstly, errors on the input signal must be considered. We account for these using a Monte Carlo procedure, whereby each input signal data bin is independently adjusted by a random offset drawn from the Normal distribution with mean zero and standard deviation the bin error. The signal is then repositioned on the SOM, and the process repeated for 1000 iterations. In this way lightcurves with poorly defined transits can cover a range of pixels on the map, covering the map regions the lightcurve is compatible with within its error. This returns a distribution of SOM pixel locations, ($x_i$,$y_i$), with $i$ the Monte Carlo iteration index.

\begin{figure}
\resizebox{\hsize}{!}{\includegraphics{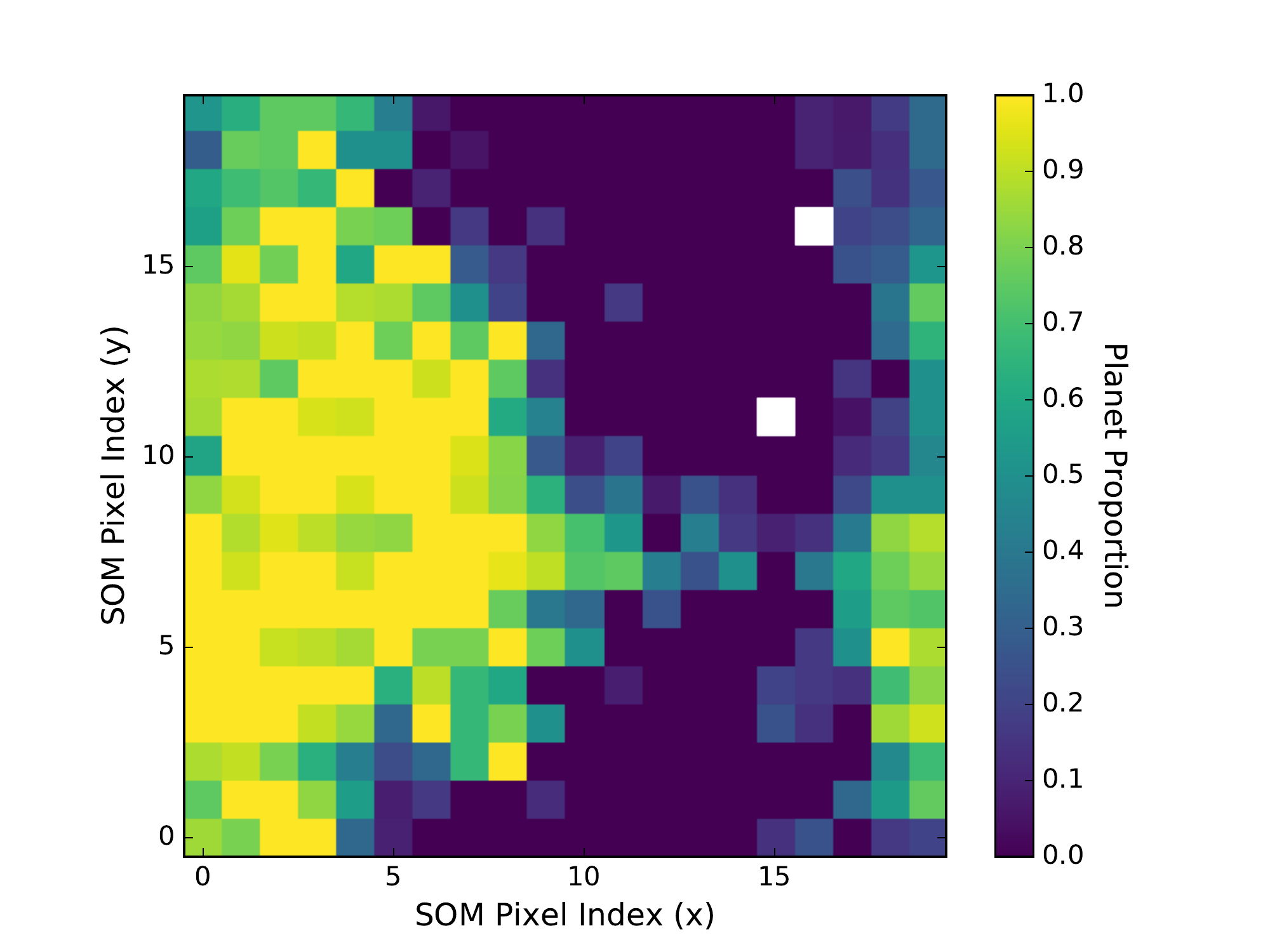}}
\caption{Proportion of dispositioned KOIs within each SOM pixel which are true planets, from the SOM in Figure \ref{figsomlockepsnrcut}. Clear grouping is seen. White pixels are those where no dispositioned KOIs are found.}
\label{figprop}
\end{figure}

The disposition of each candidate signal is then calculated as follows. For each SOM pixel, we take the proportion of already disposition signals within it that are planets, and the proportion that are false positives. Each SOM pixel in the distribution ($x_i$,$y_i$) then moves a candidate signal towards either planetary or false positive status based on the pixel's characterization (see Figure \ref{figprop}). We further weight each SOM pixel on the number of known signals within it, meaning more well characterised pixels are given increased classification power. The weights $W$ and proportion of of dispositioned signals in pixel ($x$,$y$) which are planets, $\alpha_\textrm{planet}(x,y)$ are found by

\begin{equation}
W(x,y) = \sum_o \left( x_o=x,y_o=y  \right)
\end{equation}

and

\begin{equation}
\alpha_\textrm{planet}(x,y) = \frac{\sum_{o=\textrm{planet}} \left( x_o=x,y_o=y  \right)}{W(x,y)}
\end{equation}

where $o$ is an index representing each already dispositioned object. The output statistic is then calculated by

\begin{equation}
\label{eqntheta1}
\theta_1 = \frac{\sum_i \left(  \alpha_\textrm{planet}(x_i,y_i)W(x_i,y_i)  \right)}{\sum_i\left(W(x_i,y_i)\right)}
\end{equation}

Values of $\theta_1$ above 0.5 represent planets, and those less than 0.5 false positives. The closer to unity $\theta_1$ is, the more likely a candidate is to be a planet. We stress that this is not a posterior probability (and hence cannot be used for validation of planetary candidates), although it is related. Calibration may be possible in future to make $\theta_1$ more closely resemble a probability, but the statistic would nevertheless be subject to various biases discussed in Section \ref{sectDiscuss}. The conversion of $\theta_1$ into a true posterior probability is beyond the scope of this work, as it would need to consider factors such as galactic pointing (and hence crowding) as well as the myriad other inputs to common validation codes such as \texttt{Blender} \citep{Torres:2010eob}, \texttt{PASTIS} \citep{Diaz:2014kd,Santerne:2015bb} and \texttt{vespa} \citep{Morton:2012bv}. An example of the usage of this case is given in Sections \ref{sectKepCase1} and \ref{sectK2Case1}.

\subsection{Case 2: Early in Mission Lifetime}
\label{sectCase2}
In this case, a candidate list may have been produced but large numbers of confirmed or validated planets are not yet available. This is the situation where ranking candidates may prove the most use; it is important to select the best candidates to observe for radial velocities for example, without wasting limited telescope resources.

\begin{figure}
\resizebox{\hsize}{!}{\includegraphics{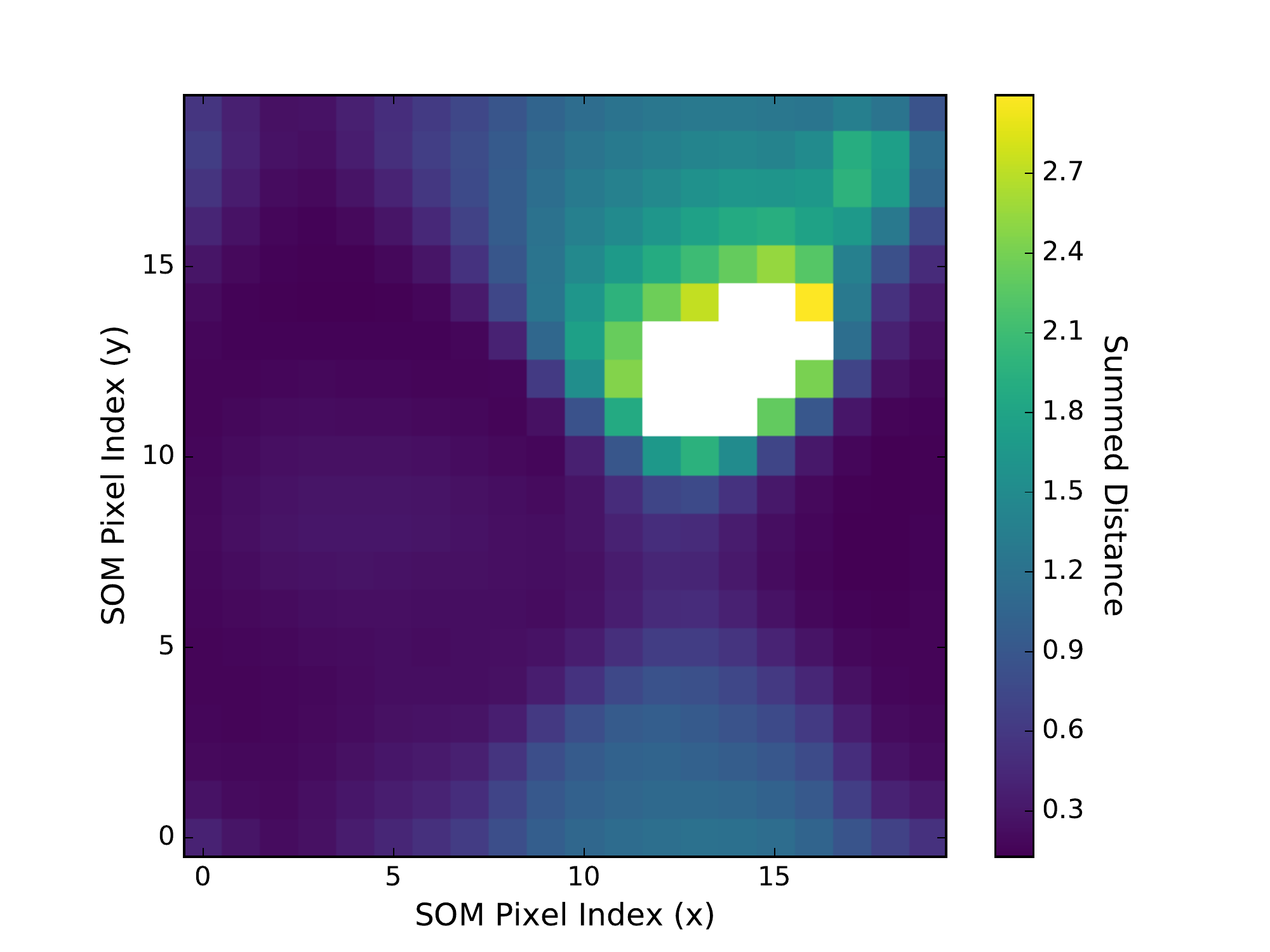}}
\caption{Summed distances between the Figure \ref{figsomlockepsnrcut} SOM pixel templates and simulated planetary signals $D_P$, as defined in Equation \ref{eqnDP}. Low distances represent a good match. Here the grouping can be seen without relying on already dispositioned KOIs. The 14 largest distances have been masked for clarity.}
\label{figdistance}
\end{figure}

We cannot use the above statistic $\theta_1$. Here we adapt $\theta_1$ to make use of simulated transit signals, created using \texttt{PASTIS} in Section \ref{sectPASTIS}. The distance between each SOM pixel and each simulation is calculated, using the sum of the squared difference between each bin, considering only the bins specifically in transit. The average distance of a pixel to each scenario's (e.g. planet, eclipsing binary, background eclipsing binary) simulation set is taken. The planet scenario distances for the \emph{Kepler} SOM of Figure \ref{figsomlockepsnrcut} are shown in Figure \ref{figdistance}. As such we calculate the average distance for each set of simulated lightcurves $D_S$, where $S$ labels the type of scenario under consideration, by

\begin{equation}
D_S(x,y) = \frac{1}{n_S}\sum_s  \sum_b \left(  T_\textrm{SOM}(x,y,b)-T_\textrm{S}(s,b) \right)^2  
\end{equation}

where $T_\textrm{SOM}(x,y,b)$ is the value of bin $b$ in SOM pixel ($x$,$y$), $T_\textrm{S}(s,b)$ is the value of bin $b$ in the simulated lightcurve $s$ of scenario $S$, and $n_S$ is the number of simulated lightcurves in scenario $S$. Next, we calculate the average distance of each SOM pixel to the planet-like and false-positive like scenarios, as

\begin{equation}
\label{eqnDP}
D_\textrm{P}(x,y) = \left<  D_{S=\textrm{planet-like}}(x,y) \right>
\end{equation}

and

\begin{equation}
D_\textrm{FP}(x,y) = \left<  D_{S=\textrm{false positive-like}}(x,y) \right>
\end{equation}

where planet-like scenarios are the P and PSB scenarios from Section \ref{sectPASTIS}. We ignore the BTP scenario as the simulated transits are extremely shallow and hence generally uninformative. False positive-like are the EB, ET and BEB scenarios, also from Section \ref{sectPASTIS}. The output statistic is then calculated as

\begin{equation}
\label{eqntheta2}
\theta_2 =    \frac{1}{n_i} \sum_i \left(  \frac{D_\textrm{FP}(x_i,y_i)}{D_\textrm{P}(x_i,y_i) +D_\textrm{FP}(x_i,y_i)}  \right)
\end{equation}

where $n_i$ is the number of Monte Carlo iterations performed. The values of $\theta_2$ have the same properties as $\theta_1$ above. Examples of this case are given in Sections \ref{sectKepCase2} and \ref{sectK2Case2}.

\subsection{Testing with PASTIS}
\label{sectPASTIStest}
\begin{figure}
\resizebox{\hsize}{!}{\includegraphics{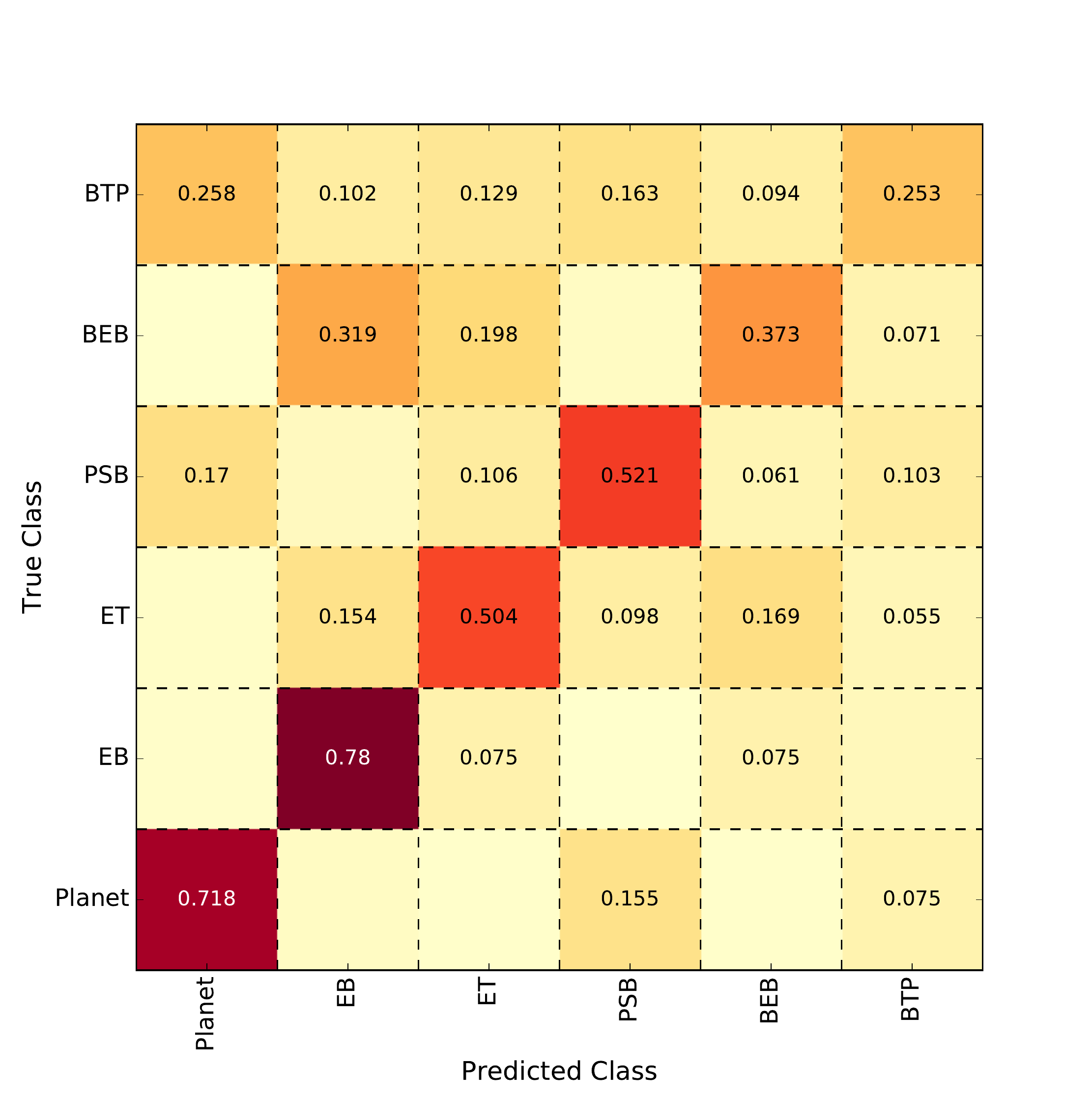}}
\caption{Confusion matrix showing the true class and predicted class for each simulated scenario described in Section \ref{sectPASTIStest}. Lightcurves are classified using the method described in Section \ref{sectCase1}. Correct classification lies along the diagonal. The proportion of lightcurves in each scenario which lie in each box is shown. Only proportions greater than 5\% are shown for clarity. Confusion is typically between the three planetary scenarios (Planet, PSB, BTP) or between the three stellar scenarios (EB, ET, BEB). Typically shallower scenarios such as BTP are less well classified.}
\label{figpastisconfmap}
\end{figure}

Given the set of simulated transit signals, it is possible to test this method for degeneracies. We perform this test by injecting the simulated signals into the \emph{Kepler} DV lightcurves, and creating binned phase folded transits as described in Section \ref{sectData}. We increase the depth of simulated transits such that each signal is at least marginally detectable. This ensures each simulation contributes information to the SOM; boosting the depth is possible as the purpose here is to test degeneracies between different scenarios rather than to find specific recovery rates. We then train a SOM with these simulations as the input data. We follow the Case 1, proportions based method of classifying to attempt to separate the injected groups. The results are shown in Figure \ref{figpastisconfmap}. Some success is found for all scenarios, but two clear groups are formed within which scenarios are degenerate. One is of scenarios where the transiting object is a planet (P, PSB, BTP), the other where the transiting object is a star (EB, ET, BEB). Some mixing between these two groups is seen towards the top right of Figure \ref{figpastisconfmap}; in this region the scenarios used have typically shallower and hence less well defined transit shapes. We conclude that the method distinguishes stellar eclipses from planetary transits successfully, but cannot exclude false positive scenarios involving planets (background transiting planets for example). This behaviour is expected, as such false positive are among the hardest to identify, and motivates the planet-like and false positive-like groups used in the Case 2 method. Furthermore transits with low SNR can be confused; this is expected to be a problem for candidates where even the binned transit has a poorly defined shape, and is allowed for in Section \ref{sectClassify} using the bin errors. The exact success rates seen in Figure \ref{figpastisconfmap} have no relevance to the success of the method on real data, as they are inherently functions of the distributions of simulated lightcurves used. The results here are only informative in so much as they highlight degeneracies between true planets and some false-positive classes found when using this method. Testing the method's results on real data is performed in Section \ref{sectKepler}.


\section{Application to Kepler}
\label{sectKepler}
\subsection{Case 1}
\label{sectKepCase1}
We take the \emph{Kepler} transit signals as prepared in Section \ref{sectData}. We found best results from training the SOM only on the higher signal-to-noise (SNR) transits, and use a cut at 30 in the SNR parameter given by the NASA Exoplanet Archive, leaving 3078 KOIs. The SOM is then trained as described in Section \ref{sectSOM}. Note that we can obtain classifications for all signals, despite only training on a subset. The locations of all the \emph{Kepler} signals on this SOM are shown in Figure \ref{figsomlockepsnrcut}, with examples of the SOM pixel templates underlying the map in Figure \ref{figtemplates}.

\begin{figure}
\resizebox{\hsize}{!}{\includegraphics{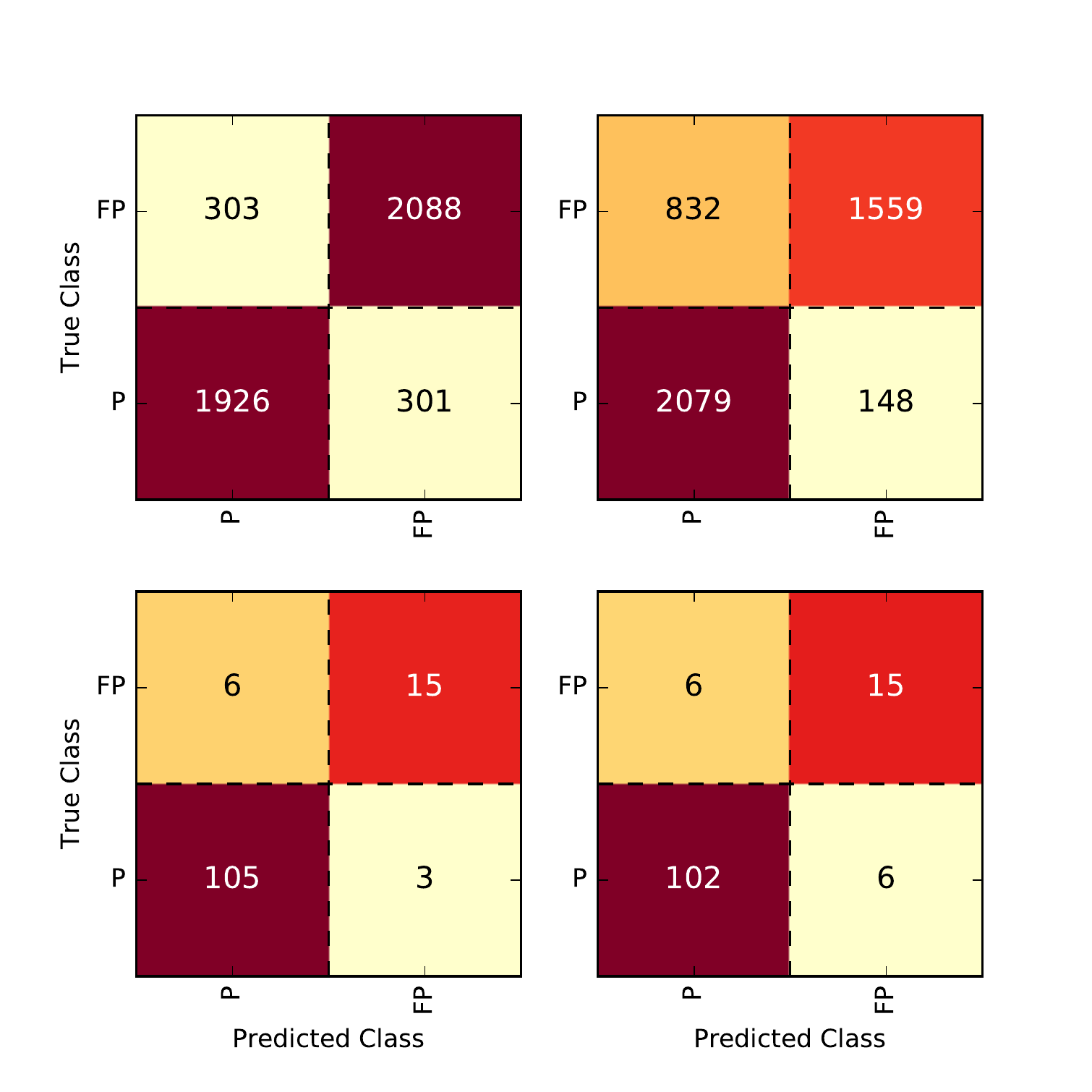}}
\caption{Confusion matrix showing the dispositions given to candidate signals, using a threshold in $\theta_1$ or $\theta_2$ of 0.5 to classify a signal. Matrices are shown for \emph{Kepler} (top) and \emph{K2} (bottom), and for Case 1, $\theta_1$ (left), and Case 2, $\theta_2$ (right). Each box shows the total number of signals classified. P represents Planets, FP False Positives.}
\label{figallconfmat}
\end{figure}

\begin{figure}
\resizebox{\hsize}{!}{\includegraphics{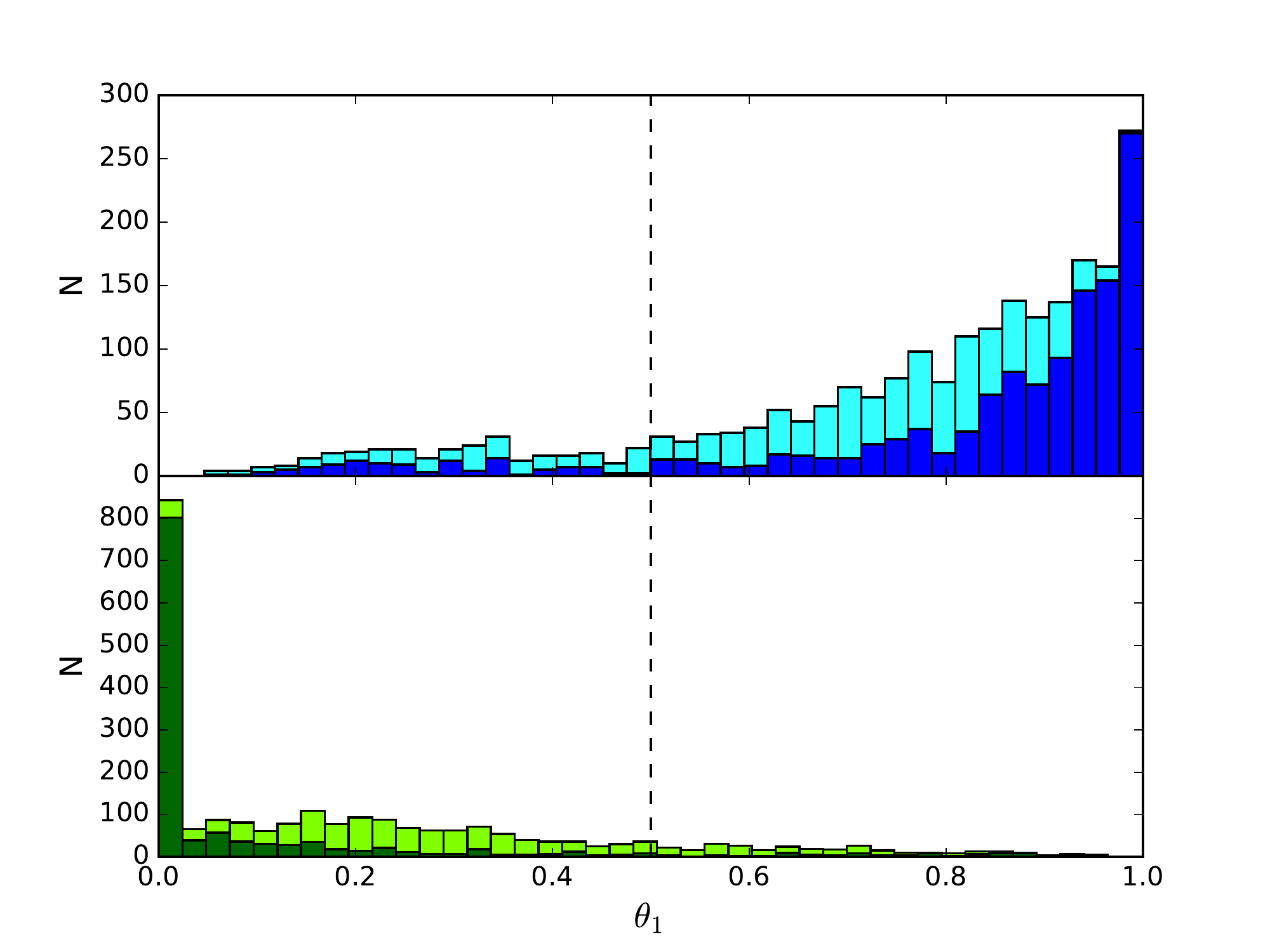}}
\caption{$\theta_1$ statistic for the \emph{Kepler} KOIs. Top: true planets. Bottom: false positives. The lighter shaded histograms show the whole sample, while darker shades represent the higher SNR KOIs.}
\label{figkephist1}
\end{figure}

We will apply both classification methods to the \emph{Kepler} data. The proportions of planets and false positives in each SOM pixel are shown in Figure \ref{figprop}. The SOM was unaware of the disposition of each signal before training. Hence we can use every input transit to test the method. We apply Equation \ref{eqntheta1} to the \emph{Kepler} signals and show a histogram of the resulting statistic in Figure \ref{figkephist1}. Values of $\theta_1$ greater than 0.5 represent planets, with the strength of the result increasing as $\theta_1$ rises. 1923 of the 2227 dispositioned planets are classified correctly, and 2093 of the 2391 false positives, making an overall success rate of 87.0\%. The success rate rises to 91.7\% when considering only the higher SNR KOIs used to train the SOM. Taking only objects with `well-determined' classification (4188 of the 4618 dispositioned KOIs, defined as $\theta_1$ greater than 0.6 or less than 0.4 for planets and false positives respectively) improves the results for all KOIs to 89.8\%, and the results for higher SNR KOIs to 92.8\%. Confusion matrices for this case and the others tested are shown in Figure \ref{figallconfmat}. We note however that the aim of this method is to rank candidates rather than classify them; Figure \ref{figkephist1} is more useful for gaining an understanding of the method's success for this purpose, through the distribution of $\theta_1$ for planets and false positives.

\subsection{Case 2}
\label{sectKepCase2}
\begin{figure}
\resizebox{\hsize}{!}{\includegraphics{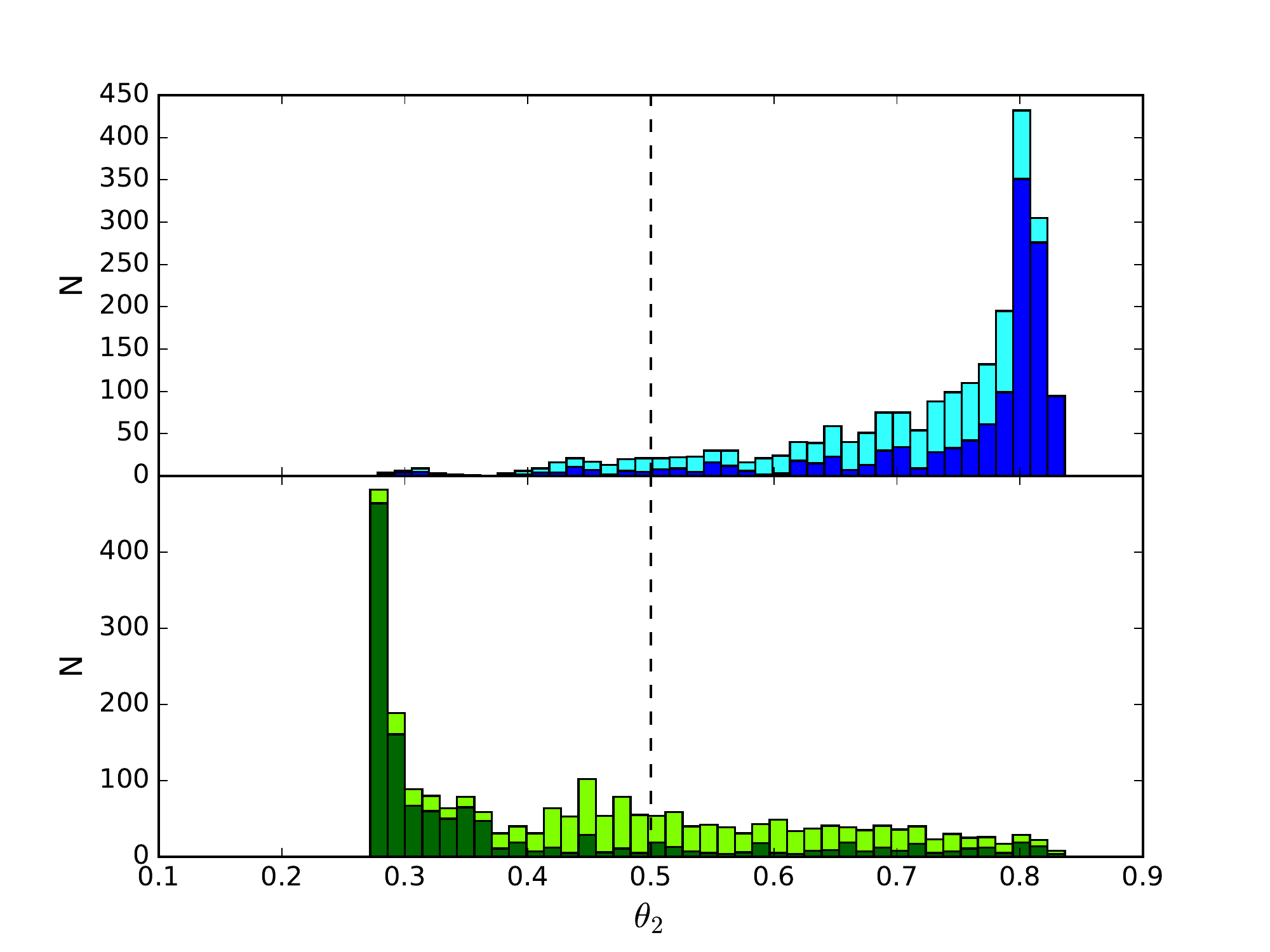}}
\caption{As Figure \ref{figkephist1} for $\theta_2$.}
\label{figkephist2}
\end{figure}

Although we expect Case 1 to work better where possible, it is interesting to compare to Case 2 (using simulated lightcurves for classification, although not for training). We use the same SOM as Section \ref{sectKepCase1} to classify the \emph{Kepler} KOIs using Equation \ref{eqntheta2}. The resulting histogram is shown in Figure \ref{figkephist2}. Note that $\theta_2$ is generally unable to fully use the parameter space between 0 and 1, because even SOM pixels heavily dominated by planets still have a finite distance to false positive simulated lightcurves and vice versa. 3638 KOIs are classified correctly, making a success rate of 78.8\% Considering only higher SNR KOIs as above increases this to 88.9\%, nearly as effective as Case 1, while retaining 3572 of the 4618 dispositioned KOIs. Considering only `well-determined' classifications increases these results to 84.6\% and 93.9\% for all KOIs and higher SNR KOIs respectively, equivalent to Case 1. A side effect of $\theta_2$ not using the full 0 to 1 space is that the distribution of $\theta_2$ is unbalanced; more planets are classified correctly than false positives, as can be seen in Figure \ref{figallconfmat} (top-right). This may be a desired outcome, if for example it is more important to maintain planets than remove false positives. The balance could be adjusted by using different thresholds of $\theta_2$, as necessary. We note that these success rates will likely change from mission to mission, but do represent the effectiveness of the method, and a good test of comparison between classification cases. It is possible that future developments may improve them, such as using different and physically motivated numbers of each simulated scenario.

\subsection{Results}
\label{sectkepresults}
The results for the \emph{Kepler} KOIs are given in Table \ref{tabkepresults}. This Table provides a ranked list of the undispositioned \emph{Kepler} candidates, which we hope to be useful to anyone considering selecting targets for followup. It can be combined with other diagnostics, such as those provided on the NASA Exoplanet Archive, or codes. It is important to be aware of the biases involved in this selection. Firstly, as demonstrated in Section \ref{sectPASTIStest}, false positives involving planetary transits are not distinguished. Secondly, as the SOM separates false positive signals primarily on V-shape, grazing planets will likely be classified as false positives. While this is regrettable, grazing planets are difficult to follow up, and are relatively few in number.

\begin{figure}
\resizebox{\hsize}{!}{\includegraphics{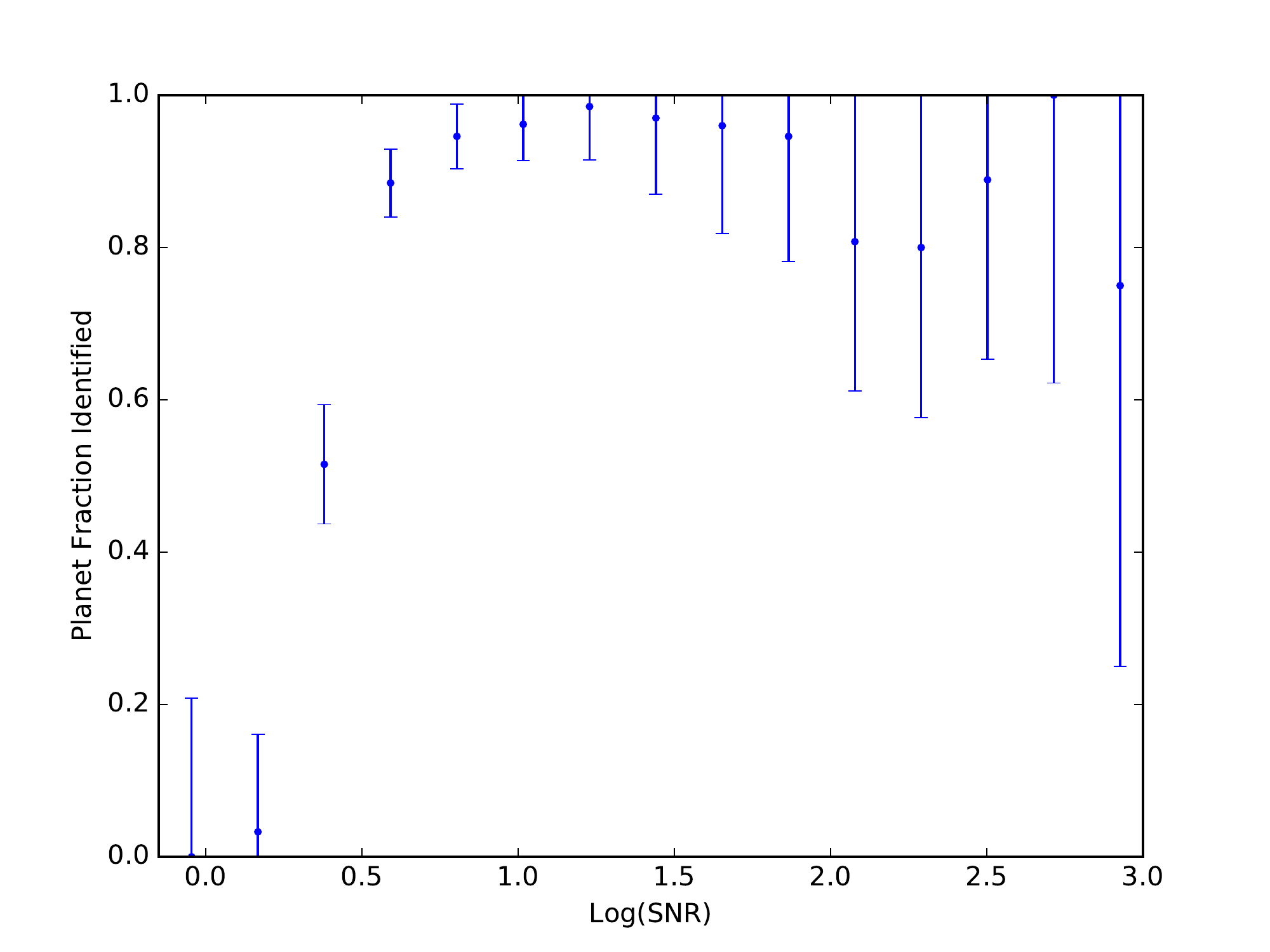}}
\caption{Fraction of successfully classified planets as a function of SNR, as defined in Section \ref{sectkepresults}. A clear drop off at low SNR is seen. Bins are spaced evenly in Log(SNR) space, and contain between 4 and 555 KOIs each. The error bars represent the Poisson counting error on the number of samples in each bin. 12 KOIs with the lowest SNR were not detected, are not shown for clarity.}
\label{figSNR}
\end{figure}

Several KOIs dispositioned as Planets are given low $\theta$ values, in both $\theta_1$ and $\theta_2$. We examined by eye the cases where $\theta_1$, the most successful method, was unable to classify planets successfully. 80\% of the 301 failure cases either showed a very low SNR signal (\mytilde 40\%), or no signal at all (\mytilde 40\%), implying that either the planet is too small to be detectable using 50 bins, or that the ephemeris provided by the archive wer erroneous. The remaining 20\% were clear transits with a V-shape, and hence are likely grazing planets as discussed above. We investigate the effects of SNR on performance by considering the ratio of planets classified correctly as a function of SNR. We estimate SNR by taking the difference between the average out-of-transit and average in-transit bins, divided by the standard deviation of the out-of-transit bins. The dependence of performance on SNR is shown in Figure \ref{figSNR}, and shows a clear decrease at low SNR. We also tested against planetary radius as given by the NASA Exoplanet Archive, and found that for low planetary radius a decrease in performance was seen (as expected given the SNR decrease which accompanies low radius). We found no other dependence on planetary radius.

\begin{table}
\caption{Output statistics for Kepler KOIs, sorted by $\theta_1$. Full table online.}
\label{tabkepresults}
\begin{tabular}{llr}
\hline
Kepler ID & $\theta_1$ & $\theta_2$ \\
\hline
005297298&1.000&0.800\\
004275191&1.000&0.699\\
003351888&1.000&0.799\\
003935914&1.000&0.699\\
008219268&1.000&0.811\\
010723750&1.000&0.800\\
009818381&1.000&0.814\\
008552719&1.000&0.797\\
005780885&1.000&0.782\\
007869917&1.000&0.795\\
010418224&1.000&0.800\\
007515679&1.000&0.799\\
\vdots&\vdots&\vdots\\
\hline
\end{tabular}
\end{table}

\section{Application to K2}
\label{sectK2}
\subsection{Case 1}
\label{sectK2Case1}
We apply the SOM to \emph{K2} similarly to \emph{Kepler}. \emph{K2} transits are binned down to 20 bins rather than 50, but we find encouragingly that the method is still effective. We do not make any SNR cuts for training the \emph{K2} SOM, due to the lower number of signals.

\begin{figure}
\resizebox{\hsize}{!}{\includegraphics{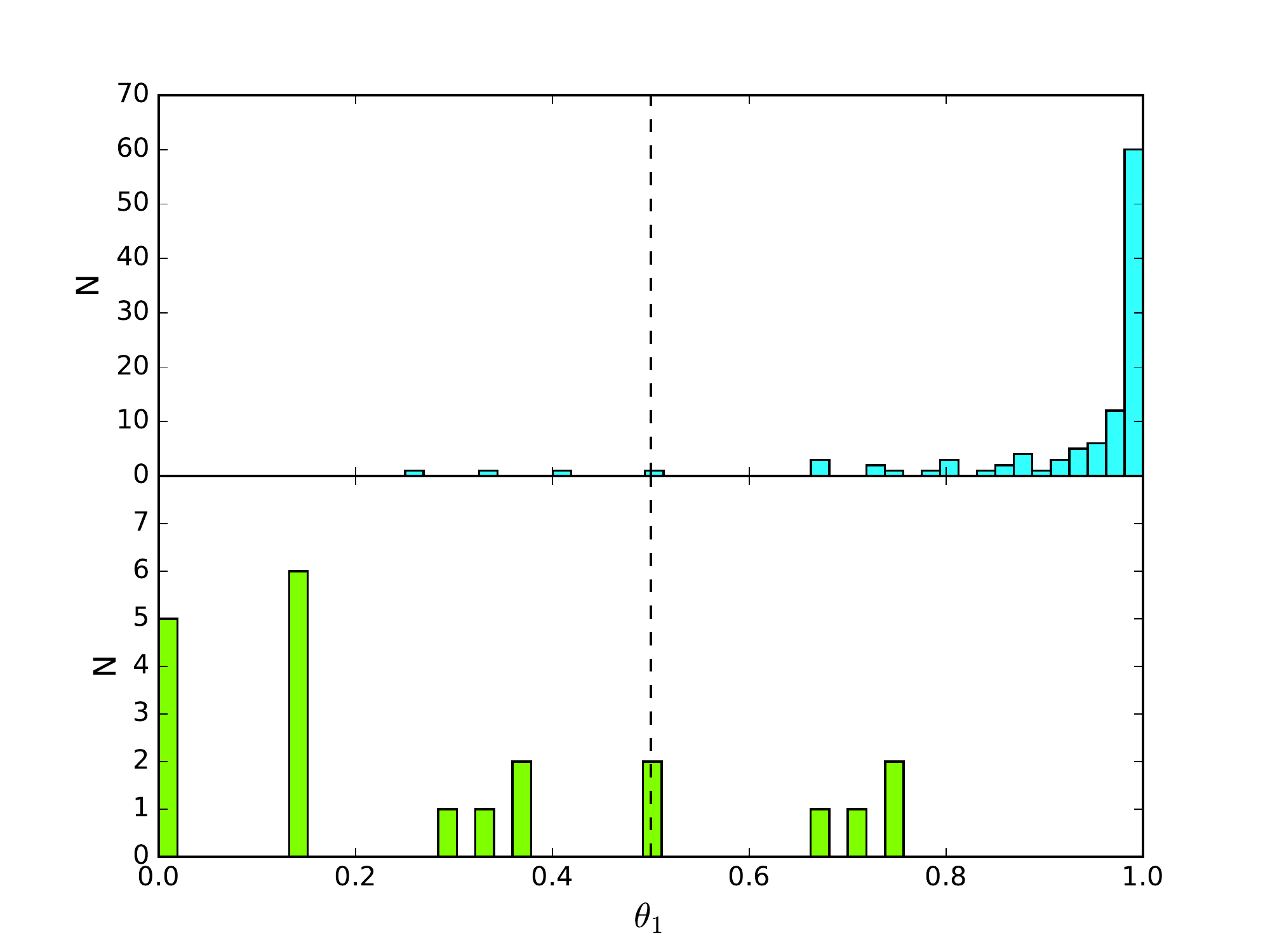}}
\caption{$\theta_1$ statistic for \emph{K2}. Top: true planets. Bottom: false positives.}
\label{figk2hist1}
\end{figure}

\begin{figure}
\resizebox{\hsize}{!}{\includegraphics{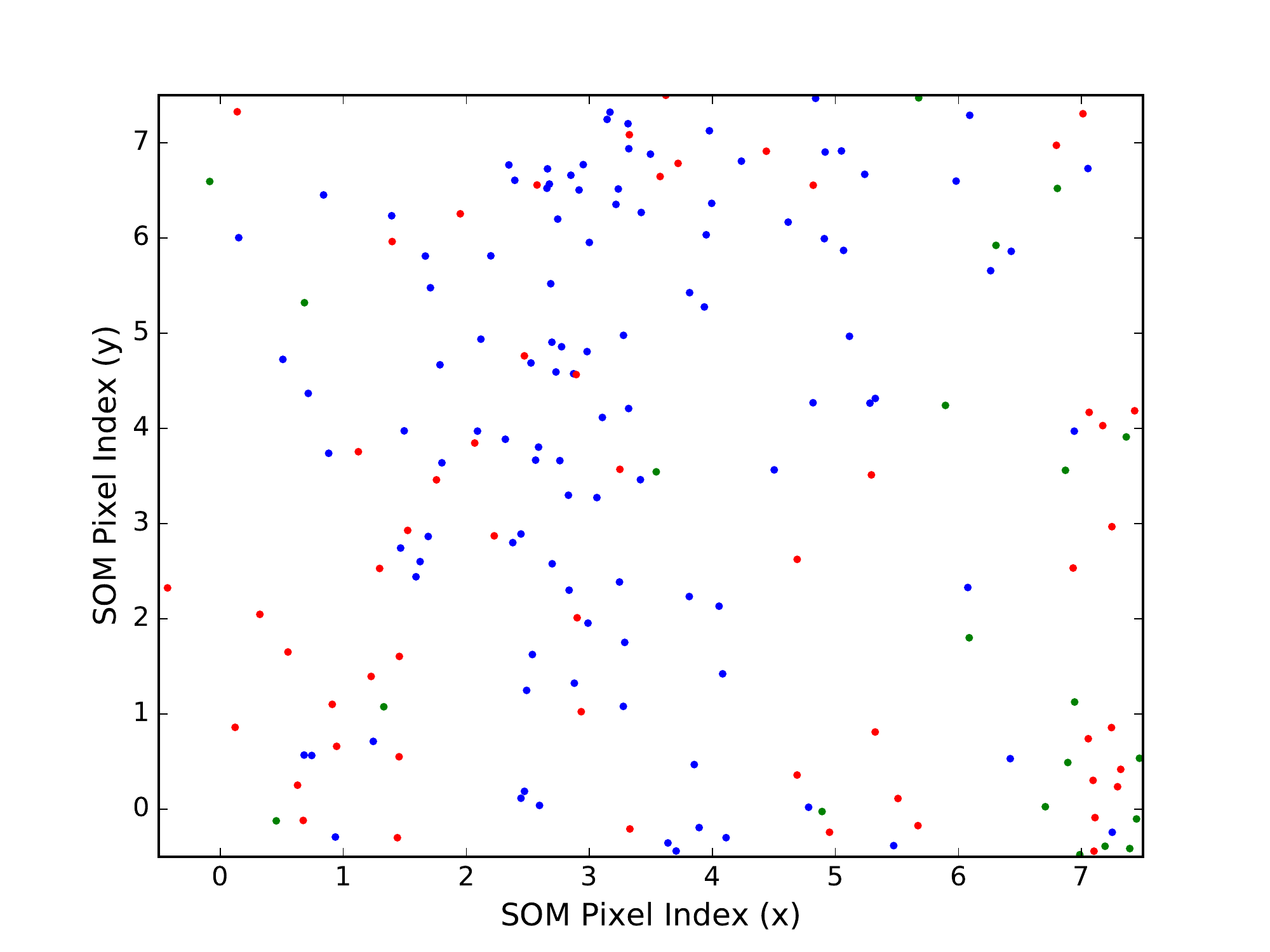}}
\caption{As Figure \ref{figsomlockepsnrcut} for the \emph{K2 sample}. Blue points represent validated planets, green validated false positives and red undispositioned candidates.}
\label{figk2somloc}
\end{figure}

\begin{figure}
\resizebox{\hsize}{!}{\includegraphics{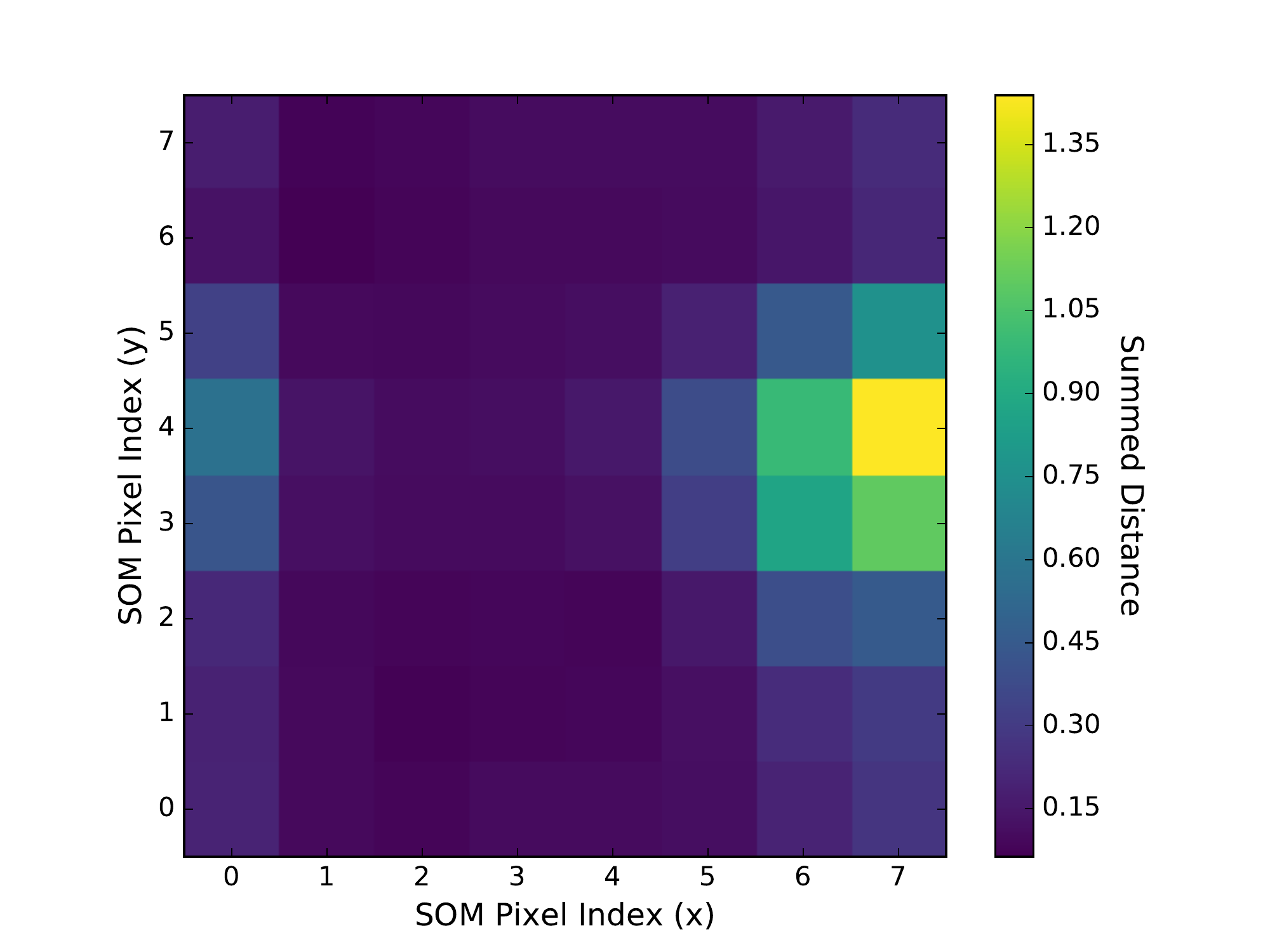}}
\caption{Summed distances between the Figure \ref{figk2somloc} SOM pixel templates and simulated planetary signals. Low distances represent a good match, and highlight the grouping seen.}
\label{figk2dist}
\end{figure}

The location of \emph{K2} signals on the trained SOM is shown in Figure \ref{figk2somloc}. The grouping is less clear due to the lower numbers, although it is apparent when considering the distances to the simulated lightcurves as described in Section \ref{sectCase2}, and shown in Figure \ref{figk2dist}. The results of attempting to use Equation \ref{eqntheta1} are shown in Figure \ref{figk2hist1}. For the planets $\theta_1$ performs well, but due to the low numbers of confirmed false positives the histogram for these is poorly populated. The success rate is reasonable, with 104 of the 108 planets classified correctly and 16 of the 21 false positives, giving a 93\% success rate overall. Given the low numbers of false positives however this is potentially spurious. Furthermore, in the case of upcoming missions such as PLATO \citep{Rauer:2014kx} or TESS \citep{Ricker:2014fy}, even this low number of dispositioned candidates will not be initially available.

\subsection{Case 2}
\label{sectK2Case2}
\begin{figure}
\resizebox{\hsize}{!}{\includegraphics{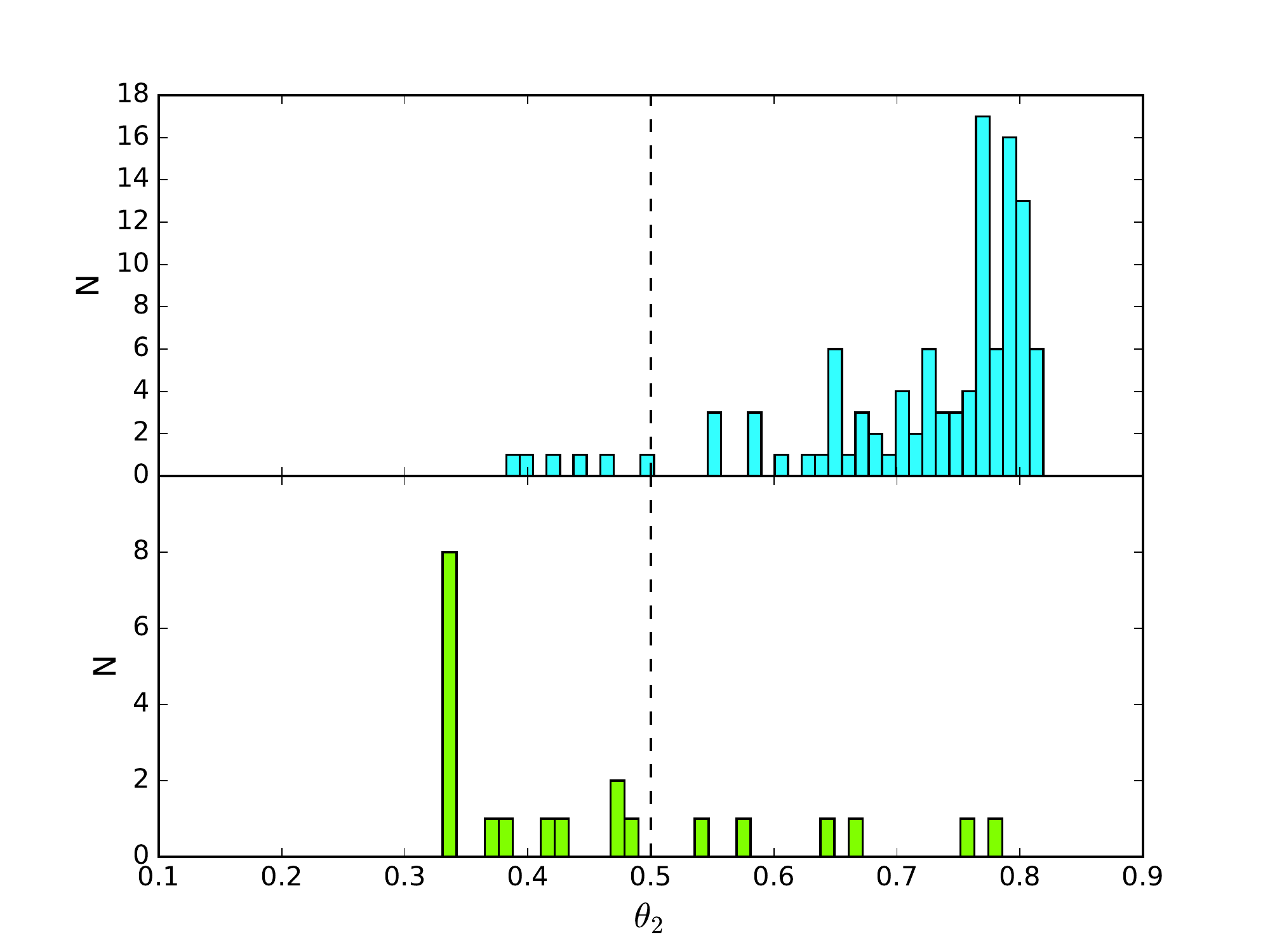}}
\caption{As Figure \ref{figk2hist1} for $\theta_2$.}
\label{figk2hist2}
\end{figure}

As such we turn to Case 2. The results of applying Equation \ref{eqntheta2} to the \emph{K2} signals are shown in Figure \ref{figk2hist2}. These have a similar success rate to Case 1, but require no already known candidates and are more robust. While still poorly populated and hence hard to test, the histogram of $\theta_2$ values for the false positives has a clearer distribution towards the low end, as desired. 102 of the 108 planets are classified correctly, and 15 of the 21 false positives, representing similar performance to Case 1. As the calculation of $\theta_2$ does not depend at all on the low number of known false positives, we believe it to be more reliable in this instance.

\subsection{Results}
The results for the \emph{K2} candidates are shown in Table \ref{tabk2results}. Users should be aware of the same caveats as highlighted in Section \ref{sectkepresults}.

\begin{table}
\caption{Output statistics for K2 objects, sorted by $\theta_1$. Full table online.}
\label{tabk2results}
\begin{tabular}{llr}
\hline
EPIC ID & $\theta_1$ & $\theta_2$ \\
\hline
201445392&1.000&0.774\\
201295312&1.000&0.770\\
201324549&1.000&0.773\\
201713348&1.000&0.755\\
201247497&1.000&0.803\\
201549860&1.000&0.781\\
201565013&1.000&0.811\\
201565013&1.000&0.794\\
201596316&1.000&0.801\\
201596316&1.000&0.754\\
201345483&1.000&0.743\\
201702477&1.000&0.797\\
201613023&1.000&0.754\\
\vdots&\vdots&\vdots\\
\hline
\end{tabular}
\end{table}


\section{Code Availability}
The code used to create and train the SOM is already part of a public package, \texttt{PyMVPA}\footnote{http://www.pymvpa.org}. To make using this method easier for readers, we make available code to classify a \emph{Kepler} or \emph{K2} lightcurve using our pre-trained SOM, along with convenience functions for users to create their own SOMs. This is available on \texttt{github}\footnote{https://github.com/DJArmstrong/TransitSOM}, along with documentation describing its use.

\section{Discussion}
\label{sectDiscuss}
SOMs have proven to be effective at separating true planetary signals from false positives, using only the shape of the candidate signal. The source of this discriminatory power is made clear by Figure \ref{figtemplates}. In short, transiting planets produce U-shaped transits, whereas most stellar eclipses are V-shaped, primarily due to the different radius ratios involved in each case. This difference is well known; it forms a key part of planetary candidate selection in most current surveys. To date however, the choice of whether a candidate is too V-shaped to continue observing has typically been made by humans, with the subjective biases and inconsistent thresholds that that implies. `How V-shaped is too V-shaped?' is a question often answered on a case by case basis. The SOM provides an opportunity to standardise, quantify and speed up this process.  We expect the SOM to be useful either as a fast pre-screen of a given candidate list, or as input to a more comprehensive autovetting code which incorporates other inputs such as secondary eclipse detections.

It is possible for true planets to give V-shaped transits. This occurs for grazing transits, where only part of the planet occults the stellar disk, as well as for near-grazing planets and short period planets observed at a very long cadence. As such, in removing V-shaped signals we are removing some true planets. This problem is common to both human selection and the SOM. While it would be preferable to maintain all planets, losing some at the expense of the majority of false positives is generally considered worthwhile given limited telescope time. Furthermore, grazing transits are difficult to model, as they present a degeneracy between radius ratio, impact parameter and inclination which is easier to separate in the full transit case.

A potential issue in our development of the SOM arises from the Kepler KOI sample. The majority of this sample have been dispositioned using validation \citep[e.g.][]{Morton:2016ka}, without separate observations of the planetary mass. This process relies on finding the false positive probability of a candidate, using its galactic pointing, local crowding, transit shape and host star parameters. As we are relying exclusively on the transit shape, one of the key validation inputs, there is a danger of bias in using the validated sample to confirm the method. We have mitigated for this by marking `confirmed' planets (those with detected masses) separately in the SOM. Confirmed planets follow the same groupings, supporting our conclusions. Furthermore, testing with \texttt{PASTIS} (Section \ref{sectPASTIStest}) successfully and independently checked the effectiveness of our method. We note that the success of the SOM demonstrates the power of the transit shape alone in the validation process, at least to the point of separating stellar eclipses from planetary transits.

In a reversal of the main goal of this work, it is possible to use the SOM to identify eclipsing binaries and triple stars. Catalogues of eclipsing binaries \citep{Armstrong:2016br,LaCourse:2015jr,Armstrong:2015bn} are useful science products of planet surveys, with the most recent catalogue for \emph{K2} using SOMs to identify the binary stars in the sample. Eclipsing binaries provide one of the only direct tests of stellar evolution models, and can even host planetary systems themselves \citep[e.g.][]{Doyle:2011ev}.

We have developed this method with the aim of separating astrophysical false positives from true planetary signals. In doing this we ignore candidates produced due to instrumental noise and apparently periodic noise patterns. These can be removed with other techniques; looking for clusters of candidates in epoch space for example. Here the SOM can also contribute. Firstly, noise-candidates do not typically show a transit-like shape. As such, they will have a large distance from even their best matching pixel on the SOM. This can be used to separate candidates. If noise-candidates are included in training the SOM, they will develop their own region on the map, one which does not resemble any simulated astrophysical lightcurve; again, this can be utilised. If all such non-matching candidates are designated false positives, the planet sample will be preserved.

\section{Conclusion}
A new method for identifying the best planetary candidates for followup has been developed, tested, and applied to the \emph{Kepler} and \emph{K2} datasets. The SOM replies only on the transit shape, and can achieve accuracies of near 90\% in distinguishing known \emph{Kepler} planets from false positives. We apply the technique to the unclassified \emph{Kepler} and \emph{K2} candidates, and hope the resulting rankings will be useful to the community.

This method adds to the developing body of techniques for automatic vetting of planetary candidates. SOMs can contribute both as a quick initial screening step and as a part of larger autovetting codes. Such codes are beginning to become available for \emph{K2} \citep{Coughlin:2016wa}, and we intend to apply this method in combination with similar techniques in the future. Autovetting is a growing field, and will become increasingly important as new missions such as \emph{TESS} and \emph{PLATO} begin to produce data. The unprecedented large data volume of these missions will require automatic techniques to maximise their effectiveness. In addition to followup efficiency, automatic techniques allow faster and more detailed studies of completion rates in planetary surveys, allowing statistical studies to be made more easily and more robustly. We expect developments in this field to progress rapidly from now on.

\section*{Acknowledgements}
We would like to thank the anonymous referee for providing useful comments on the manuscript. D..J.A. acknowledges funding from the European Union Seventh Framework programme (FP7/2007- 2013) under grant agreement No. 313014 (ETAEARTH). This publication was aided by the international team led by J. Cabrera on `Researching the Diversity of Planetary Systems' at ISSI (International Space Science Institute) in Bern. Part of this work was supported by Funda\c{c}\~ao para a Ci\^encia e a Tecnologia, FCT, (ref. UID/FIS/04434/2013 and PTDC/FIS-AST/1526/2014) through national funds and by FEDER through COMPETE2020 (ref. POCI-01-0145-FEDER-007672 and POCI-01-0145-FEDER-016886). A.S. is supported by the European Union under a Marie Curie Intra-European Fellowship for Career Development with reference FP7-PEOPLE-2013-IEF, number 627202. This paper includes data collected by the Kepler mission. Funding for the Kepler mission is provided by the NASA Science Mission directorate. The data presented in this paper were obtained from the Mikulski Archive for Space Telescopes (MAST). STScI is operated by the Association of Universities for Research in Astronomy, Inc., under NASA contract NAS5-26555. Support for MAST for non-HST data is provided by the NASA Office of Space Science via grant NNX13AC07G and by other grants and contracts.

\bibliography{papers041016}
\bibliographystyle{mn2e_fix}

\end{document}